\documentclass[twocolumn]{jpsj2} %% two-column layout
\title{Tomonaga-Luttinger liquid in quasi-one-dimensional antiferromagnet $\rm{BaCo_2V_2O_8}$ in magnetic fields}

\author{Seiichiro \textsc{SUGA}}

\inst{Department of Applied Physics, Osaka University, Suita, Osaka 565-0871}

\abst{Motivated by recent studies on the quasi-one-dimensional (1D) antiferromagnet $\rm{BaCo_2V_2O_8}$, we investigate the Tomonaga-Luttinger-liquid properties of the 1D $S=1/2$ $XXZ$ Heisenberg-Ising model in magnetic fields.  By using the Bethe ansatz solution, thermodynamic quantities and the divergence exponent of the NMR relaxation rate are calculated. 
We observe a magnetization minimum as a function of temperature $(T)$ close to the critical field. As the magnetic field approaches the critical field, the minimum temperature asymptotically approaches the universal relation in agreement with the recent results by Maeda {\it et al}. 
We observe $T$-linear specific heat below the temperature of the magnetization minimum. The field dependence of its coefficient agrees with the results based on conformal field theory. 
The field dependence of the divergence exponent of the NMR relaxation rate with decreasing temperature is obtained, indicating a change in the critical properties. The results are discussed in connection with experiments on $\rm{BaCo_2V_2O_8}$.}

\kword{1D $S=1/2$ XXZ model, magnetic field, Tomonaga-Luttinger liquid, $\rm{BaCo_2V_2O_8}$, Bethe ansatz}

\begin{document}
\maketitle
%%%%%%%%%%%%%%%%%%%%%%%%%%%%%%%%%%%%%%%%%%%%%%%%%%%%%%%%%%%%%%%%%%%%%
\section{\label{sec:level1}Introduction}
%%%%%%%%%%%%%%%%%%%%%%%%%%%%%%%%%%%%%%%%%%%%%%%%%%%%%%%%%%%%%%%%%%%%%
One-dimensional (1D) quantum spin systems have been investigated for many years. Their fascinating aspects have been revealed by theoretical and experimental studies. In particular, 1D gapped spin systems have attracted a large amount of attention \cite{rev}. There are various gapped spin systems including Haldane-gap systems, two-leg ladders, and bond-alternating systems. 
It was argued that the characteristics of each system appear conspicuously in their critical and dynamical properties above the critical magnetic field $(H_{\rm c})$ where the gap is closed and low-energy properties can be described as a Tomonaga-Luttinger liquid (TLL) \cite{chitra}. Such characteristic behavior depending on the model can be observed in, for example, the NMR relaxation rate \cite{chitra,haga}. In fact, the divergent behavior of the NMR relaxation rate with decreasing temperature was observed for $S=1$ Haldane-gap material ${\rm (CH_3)_4NNi(NO_2)_3}$ \cite{goto1,goto2} and $S=1/2$ bond-alternating chain material pentafluorophenyl nitronyl nitroxide (${\rm F_5PNN}$) \cite{izumi}. The field dependence of the divergence exponent was discussed in comparison with theoretical results \cite{sakai,suzuki1,suzuki2}. 

Recent experimental advances have allowed the study of the field dependence of thermodynamic quantities in 1D gapped spin systems in precise comparison with the theory. The specific heat of the quasi-1D $S=1$ bond-alternating antiferromagnet Ni(C$_9$H$_{24}$N$_4$)(NO$_2$)ClO$_4$ was measured in magnetic fields \cite{hagiwara}. It was found that the low-temperature specific heat is proportional to temperature $(T)$, indicative of the gapless dispersion relation. 
The field dependence of its coefficient was compared with the result based on conformal field theory (CFT) \cite{blote,affleck}. Good agreement was obtained. 
Other noticeable features of the TLL observed in its thermodynamic quantities were pointed out quite recently. The minimum of the magnetization appears as a function of temperature close to $H=H_{\rm c}$ in 1D gapped spin systems \cite{maeda}. The magnetization minimum marks the important temperature below which the TLL is preserved. It was demonstrated further that the temperature of the minimum approaches the universal relation as $H$ approaches $H_{\rm c}$ from above. 
It has been difficult to determine the temperature region for the TLL from the model Hamiltonian itself. The magnetization minimum plays an important role in determining the temperature region of the TLL not only in calculations but also in experiments. 

The quasi-1D antiferromagnet $\rm{BaCo_2V_2O_8}$ has been investigated intensively in recent years \cite{he,kimura1,kimura2,kimura3}. From the magnetization measurement, it was shown that $\rm{BaCo_2V_2O_8}$ is well described by the 1D $S=1/2$ $XXZ$ Heisenberg-Ising model \cite{kimura1}. The result has stimulated further investigations on the novel properties caused by the TLL. Since the N\'{e}el-like ground state of the model can be regarded as a charge-density-wave (CDW) insulator with quasi-long-range order from the viewpoint of the spinless fermion, it is expected that the dominant spin correlation changes from longitudinal incommensurate to transverse staggered with increasing magnetic field when $H_{\rm c}<H$. The change in the critical property can be detected via the NMR relaxation rate \cite{haga,suzuki1,suzuki2}. It was argued previously that the change in the dominant spin correlation was detected for ${\rm F_5PNN}$ via the NMR relaxation rate \cite{izumi}. However, the conclusion is not yet settled \cite{goto3}. $\rm{BaCo_2V_2O_8}$ is a potential candidate for observing the change in the dominant spin correlation in magnetic fields. 

In this paper, we investigate the TLL properties of the 1D $S=1/2$ $XXZ$ Heisenberg-Ising model in magnetic fields by using the Bethe ansatz solution. Thermodynamic quantities such as the magnetization, susceptibility, and specific heat are calculated. The temperature dependence of the magnetization is first calculated. From the temperature of the magnetization minimum, we evaluate the temperature region of the TLL in a given magnetic field close to $H=H_{\rm c}$. The results are used for discussing the TLL features obtained in other calculations. The field dependence of the coefficient of the $T$-linear specific heat is investigated. The results are compared with the field dependence of the coefficient obtained from the CFT relation $\pi/3v$ with $v$ being the excitation velocity. 
 Furthermore, the divergence exponent of the NMR relaxation rate is calculated as a function of the magnetic field. 
In \S 2, we outline the model and the method. In \S 3, we present the numerical results, which are discussed in connection with the thermodynamic and critical properties of $\rm{BaCo_2V_2O_8}$. Section 4 is a summary.

%%%%%%%%%%%%%%%%%%%%%%%%%%%%%%%%%%%%%%%%%%%%%%%%%%%%%%%%%%%%%%%%%%%%%
\section{\label{sec:level2}Model and Method}
%%%%%%%%%%%%%%%%%%%%%%%%%%%%%%%%%%%%%%%%%%%%%%%%%%%%%%%%%%%%%%%%%%%%%
We consider the 1D $S=1/2$ XXZ Heisenberg-Ising model in a magnetic field 
%%%%%%%%%%%%%%%%%%%%%%%%%%%%%%%%%%%%%%%%%%%%%%%%%%%%%%%%%%%%%%%%%%%%%%%%%%%
\begin{eqnarray}
\mathcal{H} &=& J\sum_{i} \Big(S^x_{i}S^x_{i+1} + S^y_{i}S^y_{i+1} + \Delta S^z_{i}S^z_{i+1} \Big) \nonumber\\
  &-& g\mu_{B}H \sum_{i} S_{i}^{z},  
\label{Ham1}
\end{eqnarray}
%%%%%%%%%%%%%%%%%%%%%%%%%%%%%%%%%%%%%%%%%%%%%%%%%%%%%%%%%%%%%%%%%%%%%%%%%%%
where $J>0$. We set $g\mu_{\rm B}=2$. 
By the Bethe ansatz solution, the critical field ($H_{\rm c}$) and the saturation field ($H_{\rm s}$) are given as \cite{takahashi}
$H_{\rm c} =JK(u')u'\sinh \phi/\phi$ and $H_{\rm s} =J(1+\Delta)/2$, respectively, where $K(u')$ is the complete elliptic integral with its modulus $u'$, and $\Delta=\cosh \phi$ $(\phi>0)$. The modulus $u'$ is determined by $K(u')/K(u)=\phi/\pi$ with $u'=\sqrt{1-u^2}$. 
The measured magnetization of $\rm{BaCo_2V_2O_8}$ was fitted to the magnetization curve at $T=0$ obtained from the Bethe ansatz solution \cite{kimura1}, yielding $\Delta=2.17$ and $J=30.0 \, {\rm K}$ ($k_{\rm B}=1$) \cite{com}. In this case, $H_{\rm c} \sim 0.25J$ and $H_{\rm s} \sim 1.58J$. 
It was pointed out that the ferromagnetic next-nearest-neighbor interaction $J'=-0.065J$ is required to achieve quantitatively precise agreement with the results of ESR measurements \cite{kimura2}. However, this correctional interaction does not act as frustration in the N\'{e}el-like ground state of the Hamiltonian in eq. (\ref{Ham1}). In the present calculations, therefore, we neglect its small effect. In the following, $J$ is used in units of energy. 

The formalism for the Bethe ansatz solution at a finite temperature was derived by Gaudin \cite{gaudin}. 
The basic equations for equilibrium thermodynamics are written as an infinite set of nonlinear integral equations for the pseudoenergy. To obtain accurate numerical results, we employ the technique developed for the impurity Anderson model \cite{kawakami}, where thermodynamic quantities were calculated without numerical differentiation of the free energy functional. 
To accomplish the numerical calculation, we necessarily cut off an infinite set of the nonlinear integral equations at a finite number, which is set to be 35. 
The iterative calculations for the pseudoenergies are carried out until they all converge within a relative error of $10^{-6}$. Using the obtained pseudoenergies, their derivatives are calculated in the same way within a relative error of $10^{-4}$.  

Using the thus-obtained pseudoenergies and their derivatives with respect to temperature, we calculate the temperature dependence of thermodynamic quantities in magnetic fields.

%%%%%%%%%%%%%%%%%%%%%%%%%%%%%%%%%%%%%%%%%%%%%%%%%%%%%%%%%%%%%%%%%%%%%
\section{Results and Discussion}
%%%%%%%%%%%%%%%%%%%%%%%%%%%%%%%%%%%%%%%%%%%%%%%%%%%%%%%%%%%%%%%%%%%%%
\subsection{Magnetization minimum and susceptibility}
%%%%%%%%%%%%%%%%%%%%%%%%%%%%%%%%%%%%%%%%%%%%%%%%%%%%%%
\begin{figure}[htb]
\begin{center}
\includegraphics[scale=0.6]{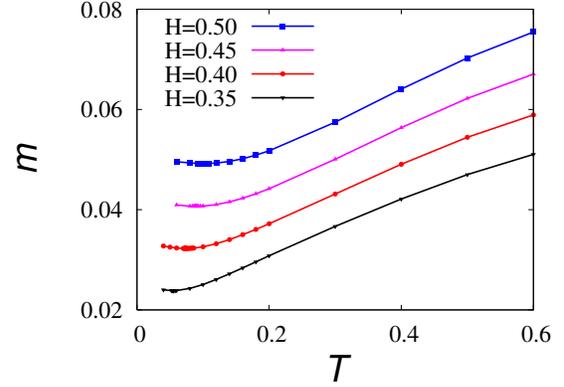}
\caption{(Color online) The magnetization in a given magnetic field as a function of temperature. }
\label{mg}
\end{center}
\end{figure}
%%%%%%%%%%%%%%%%%%%%%%%%%%%%%%%%%%%%%%%%%%%%%%%%%%%%%%
\begin{figure}[htb]
\begin{center}
\includegraphics[scale=0.6]{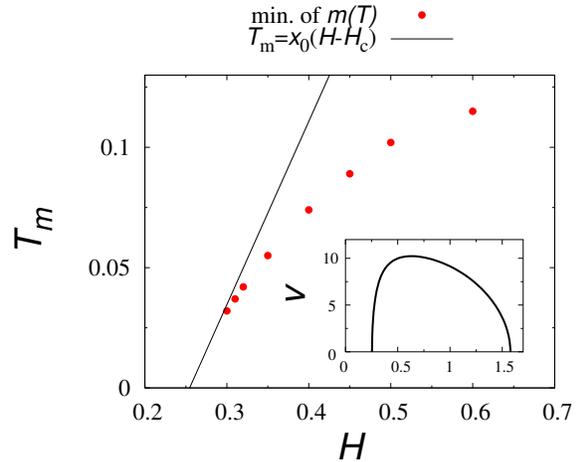}
\caption{(Color online) The temperature of the magnetization minimum as a function of the magnetic field. The solid circles represent $T_{\rm m}$ and the solid line expresses the universal relation \cite{maeda}, $T_{\rm m}=x_0(H-H_{\rm c})$, ($x_0 \sim 0.76238$). 
Inset: the velocity as a function of the magnetic field. $H_{\rm c} \sim 0.25$ and $H_{\rm s} \sim 1.58$.}
\label{tm}
\end{center}
\end{figure}
%%%%%%%%%%%%%%%%%%%%%%%%%%%%%%%%%%%%%%%%%%%%%%%%%%%%%%
We show the magnetization as a function of temperature in Fig. \ref{mg}; the magnetization has a minimum close to $H=H_{\rm c}$. The decrease in $m$ with increasing temperature is caused by the positive $\partial v/\partial H$ in the low-temperature expansion for the magnetization based on CFT \cite{maeda}. 
By using the Bethe ansatz solution at T=0, the velocity is obtained as 
$v=[d \varepsilon(\lambda)/d \lambda]/[2\pi \rho(\lambda)]|_{\lambda=B}$, where $\varepsilon(\lambda)$ and $\rho(\lambda)$ are the dressed energy and the distribution function for the rapidity $\lambda$ obtained by the following integral equations
%%%%%%%%%%%%%%%%%%%%%%%%%%%%%%%%%%%%%%%%%%%%%%%%%%%%%%%%%%%%%%%%%%%%
\begin{eqnarray}
\varepsilon(\lambda) = \varepsilon_0(\lambda) - 
   \int_{-B}^{B} \, d\lambda^{\prime} 
             a_2(\lambda-\lambda^{\prime}) \varepsilon(\lambda^{\prime}), 
\label{eqn:de}
\end{eqnarray}
%%%%%%%%%%%%%%%%%%%%%%%%%%%%%%%%%%%%%%%%%%%%%%%%%%%%%%%%%%%%%%%%%%%%
\begin{eqnarray}
\rho(\lambda) = a_1(\lambda) - \int_{-B}^{B} \, d\lambda^{\prime} 
             a_2(\lambda-\lambda^{\prime}) \rho(\lambda^{\prime}), 
\label{eqn:rho}
\end{eqnarray}
%%%%%%%%%%%%%%%%%%%%%%%%%%%%%%%%%%%%%%%%%%%%%%%%%%%%%%%%%%%%%%%%%%%%
with $\varepsilon_0(\lambda)=2H-(2\pi\sinh \phi/\phi)a_1(\lambda)$ and $a_n(\lambda)=\sinh(n\phi)/[\cosh(n\phi)-\cos \lambda]$. The cutoff $B$ is obtained by $\varepsilon(B)=0$.  
As shown in the inset of Fig. \ref{tm}, the velocity increases immediately just above $H=H_{\rm c}$ and takes a maximum around $H=0.63$. The magnetization minimum appears when $H_{\rm c}<H<0.63$ in agreement with the theoretical analysis. The temperature of the magnetization minimum ($T_{\rm m}$) vs the magnetic field is plotted by the solid circles in Fig. \ref{tm}. As the magnetic field approaches the critical field, $T_{\rm m}$ asymptotically approaches the universal relation \cite{maeda}, $T_{\rm m}=x_0(H-H_{\rm c})$, ($x_0 \sim 0.76238$). 
Since the TLL comes into existence  below $T_{\rm m}$, the result helps us to discuss the temperature region for the characteristic TLL behavior seen in thermodynamic quantities and critical exponents. 

%%%%%%%%%%%%%%%%%%%%%%%%%%%%%%%%%%%%%%%%%%%%%%%%%%%%%%
\begin{figure}[htb]
\begin{center}
\includegraphics[scale=0.6]{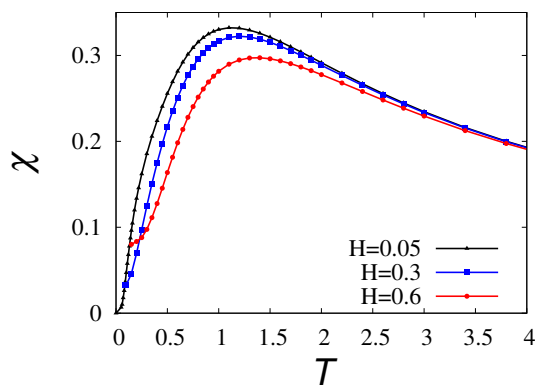}
\caption{(Color online) The temperature dependence of the susceptibilities below $(H=0.05)$, close to $(H=0.3)$, and above $(H=0.6)$ the critical field $(H_{\rm c} \sim 0.25)$.}
\label{sus}
\end{center}
\end{figure}
%%%%%%%%%%%%%%%%%%%%%%%%%%%%%%%%%%%%%%%%%%%%%%%%%%%%%%
The susceptibilities in magnetic fields below, close to, and above the critical field $H_{\rm c} \sim 0.25$ are shown in Fig. \ref{sus}. The susceptibility for $H=0.6$ approaches a finite value at $T=0$ with decreasing temperature, while the susceptibility for $H=0.05$ approaches zero, reflecting the gap. 
We compare the numerical results with the experimental ones. 
The susceptibility was  measured in magnetic fields of $H=1.0 \, {\rm T}$, $H=4.4 \, {\rm T}$, and $H=9.0 \, {\rm T}$ \cite{he}. 
Comparing the critical field $(H_{\rm c} \sim 0.25)$ and the saturation field $(H_{\rm s} \sim 1.58)$ in our calculation with the observed values, $H_{\rm c} \sim 3.9 \, {\rm T}$ and $H_{\rm s} \sim 22.7 \, {\rm T}$ \cite{kimura1}, we evaluate that the three magnetic fields in the experiment correspond to $H \sim 0.05, 0.30$, and $0.60$, respectively, in our model. 
In the experiments, the maxima of the susceptibilities were observed around $T=30 - 40 \, {\rm K}$ in the three magnetic fields. In Fig. \ref{sus}, the maxima appear at $T \sim 1.1$ for $H=0.05$,  $T \sim 1.2$ for $H=0.3$, and $T \sim 1.3$ for $H=0.6$, which correspond to $T=33 \, {\rm K}, 36 \, {\rm K}$, and $42 \, {\rm K}$, respectively. Both numerical and experimental results essentially agree with each other.

%%%%%%%%%%%%%%%%%%%%%%%%%%%%%%%%%%%%%%%%%%%%%%%%%%%%%%
\subsection{Specific heat}
%%%%%%%%%%%%%%%%%%%%%%%%%%%%%%%%%%%%%%%%%%%%%%%%%%%%%%
\begin{figure}[htb]
\begin{center}
\includegraphics[scale=0.6]{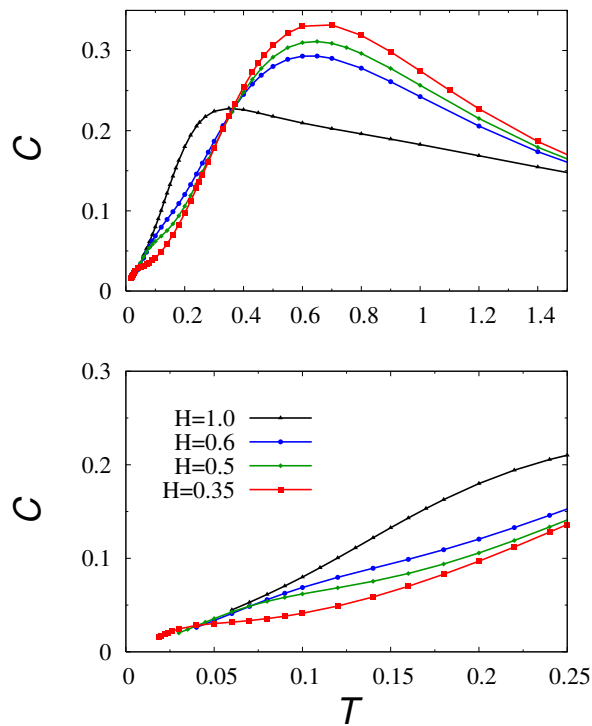}
\caption{(Color online) The temperature dependence of the specific heat in magnetic fields. The lower panel represents the specific heat in a narrow range.}
\label{spc}
\end{center}
\end{figure}
%%%%%%%%%%%%%%%%%%%%%%%%%%%%%%%%%%%%%%%%%%%%%%%%%%%%%%
\begin{figure}[htb]
\begin{center}
\includegraphics[scale=0.6]{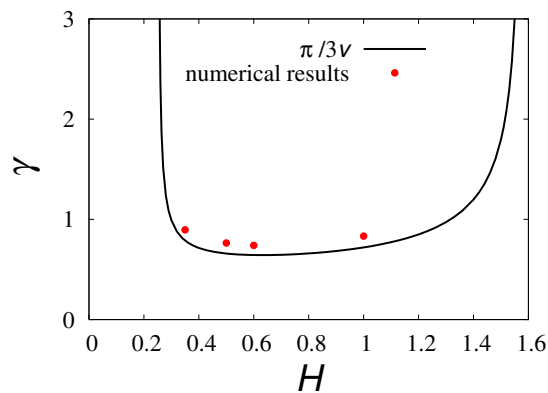}
\caption{(Color online) The coefficient of the $T$-linear specific heat in magnetic fields. The solid circles represent the results obtained numerically and the solid line represents the result based on the CFT relation.}
\label{gm}
\end{center}
\end{figure}
%%%%%%%%%%%%%%%%%%%%%%%%%%%%%%%%%%%%%%%%%%%%%%%%%%%%%%
The specific heat for $H \gtrsim H_{\rm c}$ is shown in Fig. \ref{spc}. 
When a quantum phase transition from a gapped state to a gapless state occurs in a 1D quantum many-body system, the shoulder or small peak structure may appear for the low-temperature specific heat close to the quantum critical point. This behavior was first observed for the 1D Hubbard model close to half-filling \cite{usuki,juttner}. Also for $S=1/2$ two-leg spin ladders, the low-temperature peak of the specific heat was obtained for $H>H_{\rm c}$ \cite{wang}. 
In the present system, we observe a shoulder structure at low temperatures for $H=0.35$ and $0.6$. 
It is considered that the shoulder structure is caused by the contributions of both the $T$-linear specific heat at low temperatures, which is characteristic of the TLL, and the Schottky-type specific heat. Indeed, the $T$-linear specific heat emerges below $T \sim T_{\rm m}$ and the shoulder structure appears slightly above it: $T_{\rm m}=0.05$ for $H=0.35$ and $T_{\rm m}=0.12$ for $H=0.6$. 
In Fig. \ref{gm}, we compare the coefficient of the $T$-linear specific heat obtained by the low-temperature calculations with that obtained by the CFT relation $\gamma=\pi/3v$ \cite{blote,affleck}. The former is represented by the solid circles and the latter is represented by the solid line. Both field dependences agree with each other. 
We have confirmed quantitatively that the shoulder structure of the low-temperature specific heat is caused by the TLL nature. 

Note that the low-temperature specific heat at $H=H_{\rm c}$ is proportional to $\sqrt{T}$ \cite{takahashi}, reflecting the wave-number $(q)$ dependence of the lower boundary of the excitation continuum: 
$\epsilon(q)=-H_{\rm c}+H_{\rm c}\sqrt{1+(u/u')^2 \sin^2q} \sim (H_{\rm c}/2)(u/u')^2 q^2$ for $q \sim 0$. Unfortunately, we could not reach such a low temperature region as to be able to find $\sqrt{T}$ dependence. It is a challenging issue to obtain $C \sim \sqrt{T}$ in a low temperature region at $H=H_{\rm c}$. 

We find that the shoulder structures appear at $T \lesssim 0.15$ for $H=0.35$ and at $T \lesssim 0.2$ for $H=0.6$. The magnetic fields $H=0.35$ and $0.6$ are evaluated as $H=5.3 \, {\rm T}$ and $H=8.8 \, {\rm T}$, respectively. Using $J=30.0 \, {\rm K}$, the temperature regions for the shoulder structure are evaluated as $T \lesssim 4.5 \, {\rm K}$ for $H=5.3 \, {\rm T}$ and $T \lesssim 6 \, {\rm K}$ for $H=8.8 \, {\rm T}$. 
According to experiments on the specific heat for $\rm{BaCo_2V_2O_8}$ \cite{kimura3}, no phase transition to the ordered state takes place at least at $T \gtrsim 1 \, {\rm K}$ for $H>5 \, {\rm T}$. Therefore, the $T$-linear specific heat as well as the shoulder structure are expected to be observed. By comparing the observed field dependence of $\gamma$ and the results shown in Fig. \ref{gm}, the TLL behavior of $\rm{BaCo_2V_2O_8}$ can be discussed quantitatively.

%%%%%%%%%%%%%%%%%%%%%%%%%%%%%%%%%%%%%%%%%%%%%%%%%%%%%%
\subsection{NMR relaxation rate}
%%%%%%%%%%%%%%%%%%%%%%%%%%%%%%%%%%%%%%%%%%%%%%%%%%%%%%
\begin{figure}[htb]
\begin{center}
\includegraphics[scale=0.6]{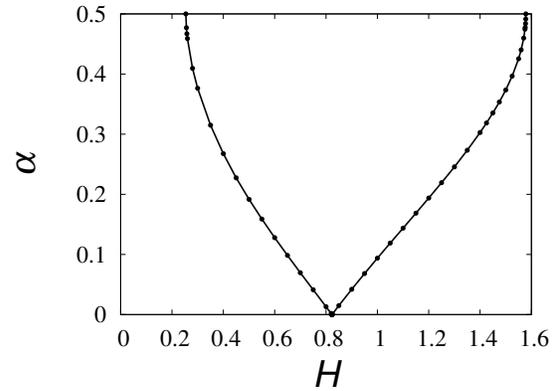}
\caption{The field dependence of the divergence exponent of the NMR relaxation rate. The V-like behavior represents the change in the critical property in magnetic fields.}
\label{alp}
\end{center}
\end{figure}
%%%%%%%%%%%%%%%%%%%%%%%%%%%%%%%%%%%%%%%%%%%%%%%%%%%%%%
We next investigate critical properties for $H_{\rm c}<H<H_{\rm s}$. 
The critical exponent of the longitudinal spin correlation function is given by $\eta^z=2[Z(B)]^2$, where 
$Z(\lambda) = 1 - (1/2\pi) \int_{-B}^{B} \, d\lambda^{\prime}  a_2(\lambda-\lambda^{\prime}) Z(\lambda^{\prime})$ \cite{bik}. 
The critical exponent $\eta^z$ takes the value of $1/2$ at $H=~H_{\rm c}$. As the magnetic field increases, $\eta^z$ increases monotonically and takes the value $2$ at $H=H_{\rm s}$. From the viewpoint of the critical properties, the gapped state for $H<H_{\rm c}$ can be regarded as the CDW insulator with quasi-long-range order, while the fully spin-polarized state for $H_{\rm s}<H$ can be regarded as the band insulator. 
Since the weak universality $\eta^z \cdot \eta^x=1$ is preserved in the TLL with $\eta^x$ being the critical exponent of the transverse spin correlation function, the dominant spin correlation changes at $\eta^z=1$ for $H_{\rm c}<H<H_{\rm s}$ \cite{haga,suzuki1,suzuki2}. 
In the TLL of 1D gapped spin systems in magnetic fields, the NMR relaxation rate ($1/T_1$) shows divergent behavior, which takes the form of $1/T_1 =A(H) \, T^{-\alpha}$. The divergence exponent $\alpha$ depends on the magnetic field and  is described in terms of the critical exponent of the dominant spin correlation function, reflecting the characteristic critical properties. The coefficient $A(H)$ depends not on temperature but only on magnetic field \cite{nmr}. 

We find that $\eta^z=1$ at $H=0.82$, which can be evaluated as $H=12.0 \, {\rm T}$. Accordingly, the incommensurate spin-density correlation is dominant with $\alpha=1-\eta^z$ for $3.9 \, {\rm T}<H<12.0 \, {\rm T}$, while the transverse staggered spin correlation is dominant with $\alpha=1-\eta^x$ for $12.0 \, {\rm T}<H<22.7 \, {\rm T}$. At $H=12.0 \, {\rm T}$, the divergent behavior of the NMR relaxation rate vanishes, because $\eta^z =\eta^x =1$. 
In Fig. \ref{alp}, we show the field dependence of $\alpha$. 
The temperature region for the TLL around $H=0.82$ $(12.0 \, {\rm T})$ can be roughly estimated as $T \lesssim 0.13$ $(3.9 \, {\rm K})$ from the extrapolation of the results shown in Fig. \ref{tm}. 
In the experiments on the specific heat, the phase transition to the ordered state was not observed until $T=0.4 \, {\rm K}$ for $H \gtrsim 10 \, {\rm T}$ \cite{kimura3}. 
Therefore, it is expected that the V-like behavior of $\alpha$ caused by the change in the critical property is observed below $3.9 \, {\rm K}$ around $H=12.0 \, {\rm T}$ for $\rm{BaCo_2V_2O_8}$.

%%%%%%%%%%%%%%%%%%%%%%%%%%%%%%%%%%%%%%%%%%%%%%%%%%%%%%%%%%%%%%%%%%%%%
\section{\label{sec:level3}Summary}
%%%%%%%%%%%%%%%%%%%%%%%%%%%%%%%%%%%%%%%%%%%%%%%%%%%%%%%%%%%%%%%%%%%%%
We have investigated the TLL properties of the 1D $S=1/2$ $XXZ$ Heisenberg-Ising model in magnetic fields. The thermodynamic quantities and the divergence exponent of the NMR relaxation rate have been calculated by using the Bethe ansatz solution. 
We have found a magnetization minimum close to $H=H_{\rm c}$. As $H$ approaches $H=H_{\rm c}$ the minimum temperature $T_{\rm m}$, below which the TLL comes into existence, approaches the universal relation asymptotically. We have observed the $T$-linear specific heat for $T \lesssim T_{\rm m}$. The field dependence of its coefficient agrees with the results based on the CFT relation. 
We have also demonstrated the V-like field dependence of the divergence exponent of $1/T_1$, which reflects the change in the critical property. 
The temperature and magnetic field regions to observe the above features in $\rm{BaCo_2V_2O_8}$ have been evaluated. 

%%%%%%%%%%%%%%%%%%%%%%%%%%%%%%%%%%%%%%%%%%%%%%%%%%%%%%%%%%%%%%%%%%%%%
\section*{Acknowledgments}
%%%%%%%%%%%%%%%%%%%%%%%%%%%%%%%%%%%%%%%%%%%%%%%%%%%%%%%%%%%%%%%%%%%%%
The author would like to thank C. Hotta, S. Kimura, and T. Suzuki for useful comments and valuable discussions. Numerical computations were carried out at the Supercomputer Center, the Institute for Solid State Physics, University of Tokyo. This work was partly supported by a Grant-in-Aid for Scientific Research from the Ministry of Education, Culture, Sports, Science, and Technology, Japan. 
%%%%%%%%%%%%%%%%%%%%%%%%%%%%%%%%%%%%%%%%%%%%%%%%%%%%%%%%%%%%%%%%%%%%%

%%%%%%%%%%%%%%%%%%%%%%%%%%%%%%%%%%%%%%%%%%%%%%%%%%%%%%


\begin{thebibliography}{00}
\bibitem{com} 
The Hamiltonian considered in ref. 15 is 
$
\mathcal{H} = J\sum_{i} \{ \epsilon (S^x_{i}S^x_{i+1} + S^y_{i}S^y_{i+1} ) + S^z_{i}S^z_{i+1} \} - g\mu_{B}H \sum_{i} S_{i}^{z}. 
$
Therefore, $J$ and $\Delta$ in the Hamiltonian in eq. (\ref{Ham1}) correspond to $\epsilon J$ and $1/\epsilon$ in ref. 15, respectively.
\end{thebibliography}
\end{document}